\documentclass[aps,twocolumn,amsmath,amssymb]{revtex4}

\usepackage{graphicx}% Include figure files
\usepackage{dcolumn}% Align table columns on decimal point
\usepackage{color}

\usepackage{amsmath}
\usepackage{amssymb}

\def\bea{\begin{eqnarray}}
\def\eea{\end{eqnarray}}
\def\ben{\begin{equation}}
\def\een{\end{equation}}
\def\benu{\begin{enumerate}}
\def\enu{\end{enumerate}}

\def\sss{\scriptscriptstyle\rm}

\def\1var{(\bx_1...\bx\N)}

\def\br{{\bf r}}

\def\bx{{x}}

\def\xc{_{\sss XC}}

\def\N{_{\sss N}}

\def\unif{^{\rm unif}}

\def\ALDA{^{\rm ALDA}}

\def\sph_int{ {\int d^3 r}}

\begin{document}
%\large\sf

%\preprint{UCI DFT GROUP: preprint FWZB07}

\title{Atoms in boxes: from confined atoms to electron-atom scattering}
\author{Meta van Faassen}
\affiliation{Department of Theoretical Chemistry, Faculty of Sciences, Vrije Universiteit Amsterdam, De Boelelaan 1083, NL-1081 HV Amsterdam, The Netherlands}

\date{\today}

\begin{abstract}
We show that both confined atoms and electron-atom scattering can be described by a unified basis set method. The central idea behind this method is to place the atom inside a hard potential sphere, enforced by a standard Slater type basis set multiplied by a cutoff factor. For confined atoms, where the wall is placed close to the atomic nucleus, we show how the energy of the highest occupied atomic orbital and the static polarizability of helium and neon atoms evolve with the confinement radius. To our knowledge, these are the first confined atom polarizability calculations that include correlation, through the use of time-dependent density-functional theory. By placing the atom in a large spherical box, with a wall outside the electron density, we obtain scattering phase shifts using a recently developed method [M. van Faassen, A. Wasserman, E. Engel, F. Zhang, and K. Burke, Phys. Rev. Lett. {\bf 99}, 043005 (2007)]. We show that the basis set method gives identical results to previously obtained phase shifts for $e$-H and $e$-He${}^{+}$ scattering.
\end{abstract}
%\pacs{31.15.Ew, 31.10.+z, 34.80.Bm, 31.25.Jf}

\maketitle

\section{Introduction}
In this paper we study the behavior of atoms under the influence of a hard spherical potential wall. We position this wall both close to the atomic nucleus, inside the range of the free atom density, and far from the nucleus, outside the range of the free atom density. 

The first of these two cases describes a confined atom. The effects of spatial confinement are important in the description of several chemistry related phenomena. Some examples are the catalytic behavior of surfaces and cages, such as zeolites, and the properties of electronic and excited states in quantum dots. For examples of confined systems see Refs.~\cite{J96,CS04} and references therein. There are several methods available to impose the boundary condition of an infinite spherical potential on the system. One of these is to explicitly impose Dirichlet boundary conditions. This method was previously used to obtain low-energy electron-atom scattering phase shifts from (time-dependent) density-functional theory ((TD)DFT)~\cite{JE05,FWEZ07,FB08}. We can also implicitly impose the boundary condition by requiring all basis functions to vanish at the wall boundary. In the classic work of Lude{\~n}a~\cite{L77,L78} a basis set for confined atoms was generated by multiplying standard Gaussian and Slater type orbitals (GTO, STO) with a cutoff factor. More recently, other basis set methods were developed. One of these optimizes the orbital exponents by fitting them to cutoff orbitals~\cite{RPJZ96} and another generates a confining potential directly from the basis functions~\cite{PKJ99}. 

In the second case, where the wall is outside the free-atom density, the bound states of the system are hardly affected by the presence of the wall. Here the purpose of the wall is to discretize the continuum. We can obtain low-energy electron-scattering properties of atoms from these discretized continuum states~\cite{FWEZ07,S06,AK84}. Low-energy electron-scattering has received a lot of interest since the discovery of Sanche and co-workers 
that low-energy electrons can cause single and double strand breaks
in DNA molecules~\cite{BCHH00}. Such large systems are hard to describe by present scattering theories, although progress has been made in the last years~\cite{TG05,TG06,TGb06}. To provide an alternative method, an electron-molecule scattering method using (TD)DFT is currently under development~\cite{FWEZ07,FB08,WBb06,WMB05}. 
Use of the TDDFT method in conjunction with discretized continuum states has proven to be an efficient method to obtain electron-atom scattering phase shifts~\cite{FWEZ07,FB08}.

In this paper we use basis functions to describe the discretized continuum states. Finite basis sets are often used to describe (discretized) continuum functions, even though they might not seem to be the most logical choice. The wish to use existing bound state methods and treat bound and continuum states on equal footing fuels the need for such basis set methods. Often, GTOs are employed~\cite{KBJ89,NP90,MS92,CMR93,CMR98}, but the use of B-spline functions is also gaining popularity (see Ref.~\cite{BCDH01} and references therein). We use the method of  Lude{\~n}a and multiply STO basis functions with a cutoff factor to impose a hard wall outside the atom. We show that such confined basis functions are not only suitable for describing confined atoms, but also to obtain low-energy electron scattering properties. We hence obtain a unified basis set method for both confined atoms and electron-atom scattering.

In the first part of this paper we introduce the basis set for confined atoms and electron-atom scattering. Next, we explain how the properties studied in this paper, the energy of the highest occupied molecular orbital (HOMO) and polarizability for confined atoms and the electron scattering phase shifts, can be obtained from (TD)DFT. In the computational details section we explain in detail how the confined STO basis functions can be made suitable for obtaining continuum states. Finally, we show results for confined He and Ne atoms, and electron scattering from H and He${}^{+}$.

\section{Theory}
In this section we explain how we place a hard wall around an atom by introducing a cutoff function in the basis set. Next, we show how a particular choice of basis set parameters leads to a basis set suitable for obtaining discretized continuum functions. Finally, we explain how the polarizability and scattering phase shifts can be obtained from time-dependent density-functional theory (TDDFT).

\subsection{Boundary condition Slater type orbitals}
Most practical solutions of the Schr{\"o}dinger equation involve the use of basis functions of some kind. To obtain the 
lowest bound states, generally a linear combination of atomic orbitals (LCAO) expansion, employing GTOs or STOs 
centered on the different atomic nuclei, is applied. GTOs are far more popular than STOs since the required integrals can be 
easily calculated analytically. A drawback of GTO basis sets is that many GTO functions are necessary to describe the cusp behavior at the nucleus. In 
practice, contracted GTO basis sets are used, which reduce the computational effort that an uncontracted set would 
require. STOs have the advantage of possessing the required cusp behavior as well as the appropriate long-range 
decay. This allows the construction of high-quality basis sets with a relatively small number of functions. STO 
basis sets can be efficiently implemented in DFT codes when a density fitting procedure is used to obtain the Coulomb 
integrals~\cite{BER73}. A type of basis set that more recently gained popularity in atomic and molecular calculations 
are basis splines, or B-splines. B-splines have obtained widespread use in atomic physics~\cite{BCDH01} because they can be 
adapted readily to the problem under investigation. This means that a B-spline based code, optimized to 
the study of bound states, can be adapted to continuum-state problems with minimal effort. Only the 
knot set, the small set of parameters defining the basis set, needs to be changed. In the next section we compare the 
continuum basis functions we develop on the basis of STOs to a typical B-spline basis. In this section we focus on 
the use of STO basis functions to describe confined atoms. 

A hard wall boundary is imposed by multiplying the basis functions with a cutoff function. The use of cutoff functions in conjunction with STOs has been first proposed by Lude{\~n}a~\cite{L77,L78}, who suggested the following form of the STOs, which we call {\em boundary} STOs (BSTOs)
\begin{eqnarray}
\phi_{nlm}(r,\theta,\phi)&=&Nf_{k}(r;R)R_{n}(r)Y_{lm}(\theta,\phi) \nonumber \\
&=&NF^{\rm BSTO}(r)Y_{lm}(\theta,\phi).
\end{eqnarray}
In this expression $N$ is a normalization factor and $n$, $l$, and $m$ are the usual quantum numbers. The $Y_{lm}(\theta,\phi)$ are spherical harmonics. The radial part $R_{n}(r)$ is the same as for ordinary STOs
\begin{equation}
R_{n}(r)=r^{n-1}e^{-\alpha r},
\end{equation}
where $\alpha$ is the orbital exponent.
The cutoff function $f_{k}(r;R)$ is defined by Lude{\~n}a as,
\begin{equation}
f_{k}(r;R)=\begin{cases}
\left(1-\frac{r}{R}\right)^{k}& r \le R\\
0& r>R\end{cases}.
\end{equation}
This function becomes zero at a sphere of radius $R$, fixing the boundary condition at $R$. For $k\ge 2$ the first derivative of a BSTO is also zero at $R$, and the function and its derivative are continuous across $R$. We choose $k=1$, this choice has been found to lead to the most accurate results for confined atoms~\cite{PKJ99}.Ê For $k=1$ the first derivative has a jump discontinuity at $R$, which obviously introduces problems when calculating first and second derivatives numerically. Therefore we take the analytic expressions of all necessary derivatives, avoiding numerical problems. 

%When cartesian STOs are used, we can, with little modification of the cutoff function, calculate the properties of an atom confined in a
%cubical box. In this paper we only consider spherical boxes with a central atom, one of the reasons being that our scattering equations 
%are defined for this particular case.

\subsection{A BSTO set for obtaining pseudo continuum states}\label{sec:bsto}
The BSTOs turn out not only to be suitable for calculating properties of confined atoms, but also to calculate discretized continuum states and scattering phase shifts. Ordinary STO or GTOs are in general not suitable for obtaining scattering states because of their exponential decay, which is notably different from the oscillating behavior of continuum states. At a first glance, the same seems to be true for BSTO basis functions, but they turn out to be quite suitable for the job at hand. The main reason for this difference is the use of the cutoff function. A normal 
STO (or GTO) will always become exponentially small at large radii and therefore is not able to accurately describe 
an oscillating continuum state. We can choose the exponential coefficients of our BSTO basis such that some 
of our basis functions are not exponentially small at the radius of the confining sphere $R$. The cutoff function then 
ensures that these basis functions are zero exactly at $R$, but nonzero before. 

In this section we explain how we obtain the parameters that define our BSTO basis. We show why this choice of parameters is especially suitable for scattering calculations and show their similarity to B-spline functions, which have already proven their merit in scattering calculations~\cite{BCDH01}. 

Our BSTO basis consists only of functions with $n\ge 2$. These functions have maxima located at some value of $r$, $r_{\rm max}$. At $r_{\rm max}$ we have
\begin{eqnarray}
\left.\frac{dF^{\rm BSTO}(r)}{dr}\right|_{r=r_{\rm max}}=0=\frac{N}{R}e^{-\alpha r_{\rm max}}\left[(n-1)Rr_{\rm max}^{n-2}\right. \nonumber\\
\left. -(n+\alpha R)r_{\rm max}^{n-1}+\alpha r_{\rm max}^{n}\right].
\end{eqnarray}
From this equation we obtain a relation between the unknown parameters $n$ and $\alpha$ and the known parameters $R$ and $r_{\rm max}$. We still have two unknowns at this point. We make the choice to choose $n$ and determine $\alpha$ for a given $n$ from 
\begin{equation}
\alpha=\frac{r_{\rm max}+n(r_{\rm max}-R)}{r_{\rm max}(r_{\rm max}-R)}
\end{equation}
We now construct the BSTO basis by choosing a value of $r_{\rm max}$ and $n$ for each function. In practice we spread out the values for $r_{\rm max}$ evenly between 0 and $R$. The value of $n$ determines the diffuseness of the BSTO functions.

\begin{figure}[tbp]
\includegraphics[width=8.5cm]{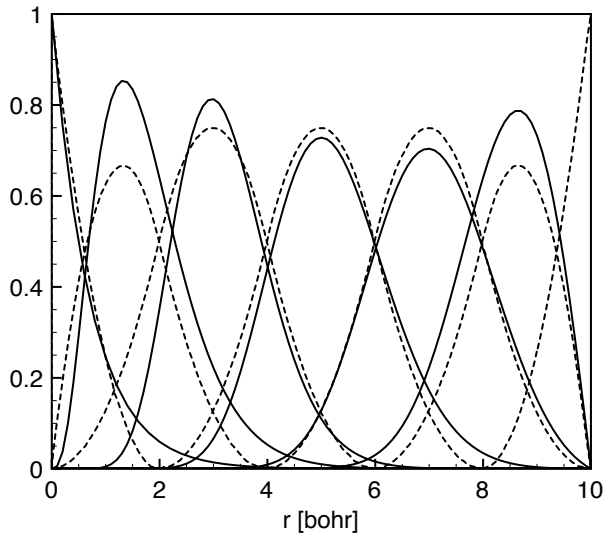}
\caption{Full lines: BSTO basis consisting of 5 functions with $n_{\rm start}=3$ and $R=10$ together with the function $(1-r/R)e^{-1.3 r}$. The maxima are chosen to coincide with the maxima of the B-spline basis functions. Dashed lines: B-spline basis set of order $k=3$ with knot sequence \{0,0,0,2,4,6,8,10,10,10\}.}
\label{f:figure1}
\end{figure}
Our choice of the BSTO parameters is best illustrated with an example. In Fig.~\ref{f:figure1} we show a typical B-spline basis set with 7 functions of order 3 (dashed lines). B-spline basis functions have already been used successfully to obtain continuum properties of atoms and molecules~\cite{BCDH01}. In the same figure we plot a BSTO basis set, where the $r_{\rm max}$ are chose to coincide with the B-spline maxima. The similarity between the BSTO and B-spline basis set is clear. An obvious difference is that the B-spline functions are only defined between the knots and are zero outside, while the BSTO functions are nonzero in the entire range from 0 to $R$. Based on the similarities between our BSTO basis set and the B-spline functions we expect the BSTO basis set to be successful in describing continuum states just as the B-spline basis.

\subsection{Properties from TDDFT}
In this section we explain how to obtain bound-state and scattering properties from TDDFT~\cite{RG84,L01,EFB09} when a hard wall is imposed. The properties considered are the polarizability and electron-scattering phase shift. 
 
The TDDFT equations are usually solved with the linear response regime. The key quantity in this regime is the frequency dependent first order induced density
 \begin{equation}
 \delta\rho(\br,\omega)=\int\chi_s(\br,\br',\omega)\delta v_s(\br',\omega) d\br'.
\end{equation}
The induced dipole moment, which is directly related to the polarizability, can be obtained from the linear response density as
 \begin{equation}
 \delta\mu(\omega)=-\int\br\delta\rho(\br,\omega)d\br
\end{equation}
and is related to the polarizability by
\begin{equation}
\mu_i(\omega)=\mu_i^{(0)}+\sum_j\alpha_{ij}E_{{\rm ext},j}(\omega).
\end{equation}
To obtain the induced density we need to know the induced Kohn-Sham (KS) potential, $\delta v_s(\br,\omega)$, and the KS response function $\chi_{s}$. The induced KS potential consists of the induced Hartree potential and an exchange-correlation (xc) potential. The induced xc-potential is obtained from the xc-kernel, $f\xc(\br,\br',\omega)$, for which the simplest choice is the adiabatic local density approximation (ALDA),
\begin{eqnarray}
&&f_{{\sss XC}\sigma\tau}\ALDA[\rho_0](\br,\br',t,t') = \nonumber\\
&&\delta^{(3)}(\br-\br')\, \delta(t-t')\,
\frac{d^2 e\xc\unif}{d\rho_\sigma d\rho_{\tau}}
\bigg|_{\rho = \rho_0},
\end{eqnarray}
where $\sigma$ and $\tau$ are spin indices. The density-density response function, $\chi_s$
, is obtained from
\begin{eqnarray}\label{e:response}
\chi_s(\br,\br',\omega)= 
\sum(f_i-f_{a})
\frac{ \phi_i^*({\bf r})\phi_{a}({\bf r}) \phi_{a}^*({\bf r}')\phi_{i}({\bf r}')}
{ (\varepsilon_i-\varepsilon_{a})+\omega+i\eta }.
\end{eqnarray}
 In this expression, $i$ and $a$ indicate occupied and virtual states respectively, the $f_i$ are the occupation numbers, $\phi_i(\br)$ are the KS orbitals, and $\epsilon_i$ the corresponding KS orbital energies. The response function is completely determined by ground-state DFT orbitals and orbital energies. All these equations are valid for finite systems and can still be applied when the density and orbitals are confined by a hard wall potential.

To obtain electron scattering phase shifts we bind an extra electron to the $N$-electron system, do a DFT calculation on this $(N+1)$-electron system, and subsequently excite this electron into the continuum using TDDFT~\cite{WB05}, or, equivalently, excite the electron into a discretized continuum state. We obtain the transition energies by solving the linear response equations as derived by Casida~\cite{C95}. These equation can be written in the form of a
Hermitian eigenvalue equation:
\begin{equation}
{\bf{\Omega }}^{S/T} {\bf{F}}_i^{S/T}  = \omega _i ^2 {\bf{F}}_i^{S/T}, 
\end{equation}
where the $\omega_i$ are the excitation energies and $S$ and $T$ stand
for singlet and triplet respectively. The ${\bf{\Omega}}$ matrices
are given by,
\begin{eqnarray}\label{eq:omegas}
\Omega _{ia,jb}^S  &=& \delta _{ab} \delta _{ij} ( \varepsilon_a - \varepsilon_i)^2 \\
&+& 2\sqrt {\varepsilon_a - \varepsilon_i} \left( K_{ia,jb}^{ \uparrow  \uparrow }  
+ K_{ia,jb}^{ \uparrow  \downarrow }  \right)\sqrt {\varepsilon_b - \varepsilon_j} \nonumber
\end{eqnarray}
\begin{eqnarray}\label{eq:omegat}
\Omega _{ia,jb}^T  &=& \delta _{ab} \delta _{ij} ( \varepsilon_a - \varepsilon_i )^2  \\
&+& 2\sqrt {\varepsilon_a - \varepsilon_i} \left( K_{ia,jb}^{ \uparrow  \uparrow }  
- K_{ia,jb}^{ \uparrow  \downarrow } \right)\sqrt {\varepsilon_b - \varepsilon_j}, \nonumber
\end{eqnarray}
where the $\varepsilon_{i}$ and $\varepsilon_{j}$ are occupied KS
orbitals and $\varepsilon_{a}$ and $\varepsilon_{b}$ are unoccupied KS
orbitals. The coupling matrix ${\bf K}$ is given by,
\begin{eqnarray}
 K_{ia,jb}^{\sigma \tau } ( \omega ) &=& \int \!\! \int \phi _{i\sigma } 
( {\bf r} ) \phi _{a\sigma } ( {\bf r} )
\left[ \frac{1}{ [ {\bf r} - {\bf{r'}}] } \right. \\
&& \left. + f_{\rm{xc}}^{\sigma \tau } 
( {\bf r},{\bf{r'}},\omega) \right] \phi _{b\tau } ( {\bf{r'}} )
\phi _{j\tau } ( {\bf{r'}} ) \, d{\bf r}d{\bf{r'}}, \nonumber
 \end{eqnarray}
 where $f_{\rm{xc}}^{\sigma \tau } $ is the
unknown exchange-correlation (xc) kernel, for which we use the ALDA kernel. Once the singlet and triplet transition energies are obtained, the phase shifts $\delta_{nl}$ for a short-ranged potential can be obtained from~\cite{FB08}
\begin{equation}\label{e:srphase}
\tan\delta_{nl}=-\frac{j_l(k_nR)}{n_l(k_nR)},
\end{equation}
where $j_{l}$ and $n_{l}$ are the spherical Bessel and Neumann functions, $R$ the hard wall radius, and $k_n=\sqrt{2(\omega_n-I)}$, where $I$ is the ionization energy. For a long-ranged potential we have~\cite{FB08}
\begin{equation}\label{e:lrphase}
\tan\delta_{nl}=-\frac{F_l(\eta,k_{n}R)}{G_l(\eta,k_{n}R)},
\end{equation}
where $F_{l}$ and $G_{l}$ are the regular and irregular Coulomb functions and $\eta$ is a Coulomb parameter.

\section{Computational details}
\subsection{Confined atoms}
In this paper we use the ADF program package~\cite{ADF07,VBBG01} for all confined atom calculations. We use the standard ADF even-tempered ET-pVQZ basis set to generate confined BSTO functions. The basis set used to fit the density~\cite{BER73} consists of a large number of ordinary STOs, instead of BSTOs. Because of the large number of fit functions included, the confined density is in general well described by ordinary STOs. When the confinement radius becomes very small, the diffuse fit functions can lead to numerical instabilities and the most diffuse functions are either removed or replaced by more contracted functions to accurately describe the density of the confined atoms.

The ground-state xc-potential we use for the confined atom calculations is the statistical average of orbitals potential (SAOP). The SAOP potential is asymptotically correct and has proven to give good HOMO orbital energies and response properties~\cite{SGGB00}. We note that for confined atoms it is not possible to use the near exact He~\cite{UG94} and Ne~\cite{ARU98} potentials since these potentials corresponds to the density of the free, unconfined, atoms. The accurate potentials have a long ranged asymptotic tail, while the exact potential corresponding to the confined atoms is zero at the wall. The exact potentials are therefore only asymptotically correct when we do not impose a wall.  

To obtain the TDDFT polarizabilities we use the ALDA kernel.

\subsection{Scattering phase shifts}
For the BSTO phase shift calculations we use the ADF program package~\cite{ADF07,VBBG01}. A BSTO adapted standard ADF ET-pVQZ basis is augmented with BSTO functions that span the continuum as explained in section~\ref{sec:bsto}. As we mentioned, the choice of $n$ and $r_{\rm max}$ is, in principle, completely free. We let the user choose a value $n_{\rm start}\ge 2$ and then determine the values of all $M$ basis functions $n_{i}$ ($i=1,2,\ldots,M$) as,
\begin{equation}
n_{i}=\left\lfloor n_{\rm start}+(i-1)\frac{50-n_{\rm start}}{M} \right\rceil
\end{equation}
where the brackets, $\lfloor\rceil$, indicate that we round to the nearest integer of the expression on the right.
The choice of the $n_{M}$ value to always be smaller than 50 is based purely on the limits of the ADF program. The choice of letting $n_{i}$ increase steadily, generates functions of more even width and makes sure that the value of $\alpha$ does not become very small, which is numerically desirable. The ``width'' of the BSTO functions can be modified by the choice of $n_{\rm start}$. Linear dependency problem for larger number of functions $M$ can often be solved by choosing a larger value of $n_{\rm start}$. 

We spread the maxima equally in the space between 2 bohr and $R$. The reason to start at 2 bohr is that the ET-pVQZ basis already spans the region close to the nucleus completely. Adding more basis functions in that region only leads to linear dependencies in the basis set. For the atoms studied in this paper, H, He, and Ne,  2 bohr turns out to be a suitable value. 

Since we place the wall outside of the short-range part of the potential, we can, in principle, use the available near exact ground state potentials for the $(N+1)$-electron systems, H${}^{-}$ and He. The error we make in the potential at the radius $R$ is exponentially small if we choose $R$ large. In practice we only use the near exact potential for He~\cite{UG94}. In case of H${}^-$ ($e$-H${}^-$ scattering) the near exact potential is not available at enough grid points to obtain accurate results. In Ref.~\cite{FWEZ07} it was shown that for this system the EXX potential is very close to the exact potential. Therefore we use the EXX potential as input for our calculations. We apply the EXX potential as a fixed input potential, even though the ADF code has the option to perform Hartree-Fock exchange-only and hybrid potential calculations. The reason is that our BSTO method is, as yet, not implemented for the case of these orbital dependent potentials. The orbital dependence of the potentials is precisely the reason that implementation of the BSTO basis set for these potentials is considerably involved and left for future applications.

We compare the BSTO results with results obtained with a well-established fully numerical
spherical DFT code, which includes the optimized effective potential method (OEP) and is supplemented by the option to insert a hard-wall at a distance
$R$ from the origin~\cite{JE05}.
This program is basis-set independent, works with a radial grid, and
both the energies and the potentials are optimized in a
self-consistent way. The TDDFT excitation energies are calculated with a supplemental code that explicitly solves the 
radial TDDFT equations~\cite{FWEZ07}.

We use the ALDA kernel for all TDDFT calculations.

\section{Results}
\subsection{Confined atoms: He and Ne}
In this section we show results for the HOMO  energy and the static polarizability as a function of the box radius for two many-electron atoms, He and Ne.
\begin{figure}[tbp]
\includegraphics[width=8.5cm]{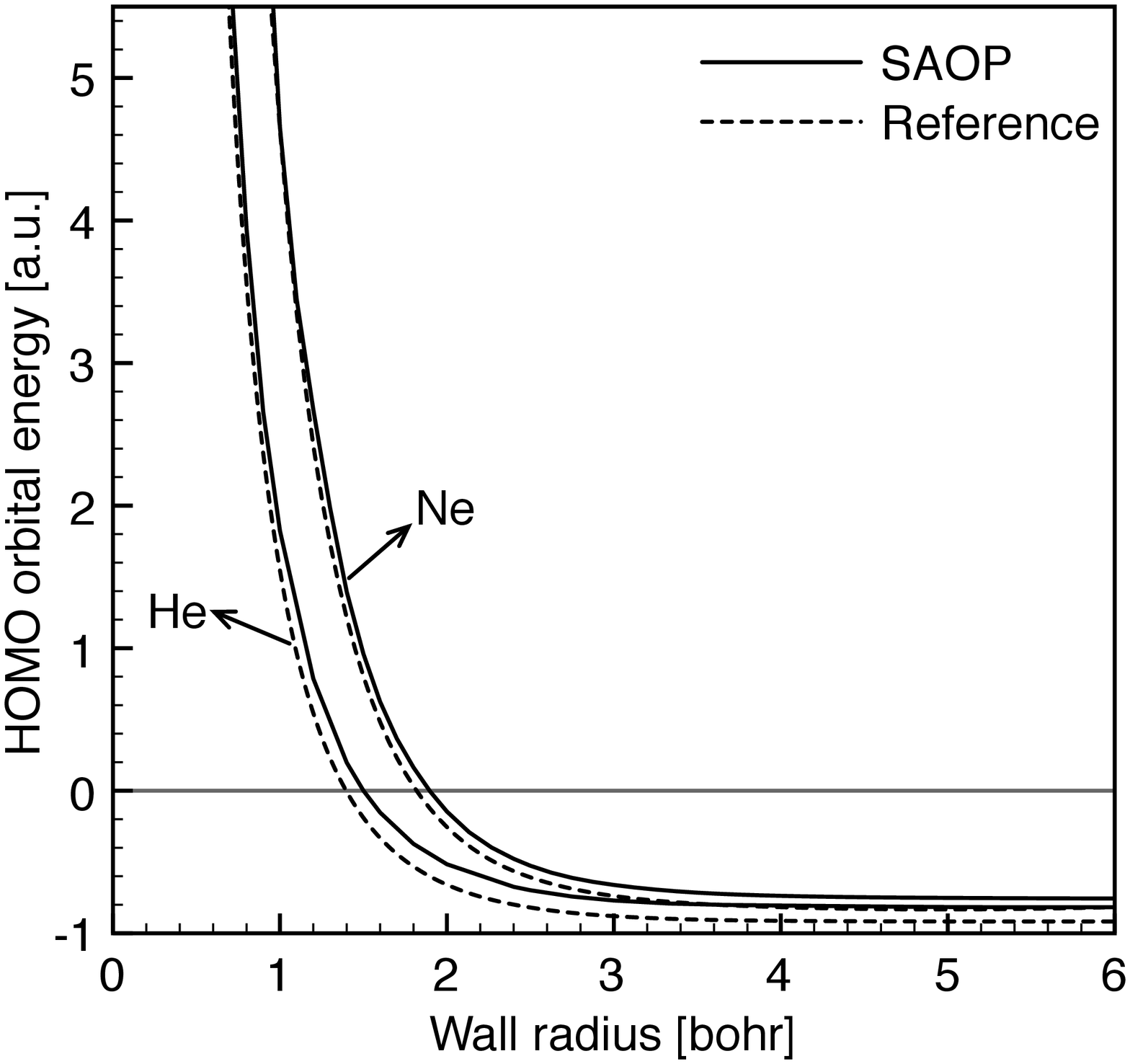}
\caption{\label{f:figure2} HOMO orbital energies of He and Ne as a function of the wall radii, compared with HF results from Ref.~\cite{L78}.}
\end{figure}
In Fig.~\ref{f:figure2} we show the change in HOMO energy with the wall radius for both He and Ne. The HOMO KS orbital decays exponentially and the wall will therefore have some influence up to $\infty$ and the free atom value is approached asymptotically. From the figure it is clear that the HOMO energies are already very close to the corresponding free atom values around 4 bohr. We obtain 95\% of the free atom He HOMO energy at a radius of 3.06 bohr. For Ne, the radius where 95\% of the HOMO energy is obtained is at 3.50 bohr. 

\begin{figure}[tbp]
\includegraphics[width=8.5cm]{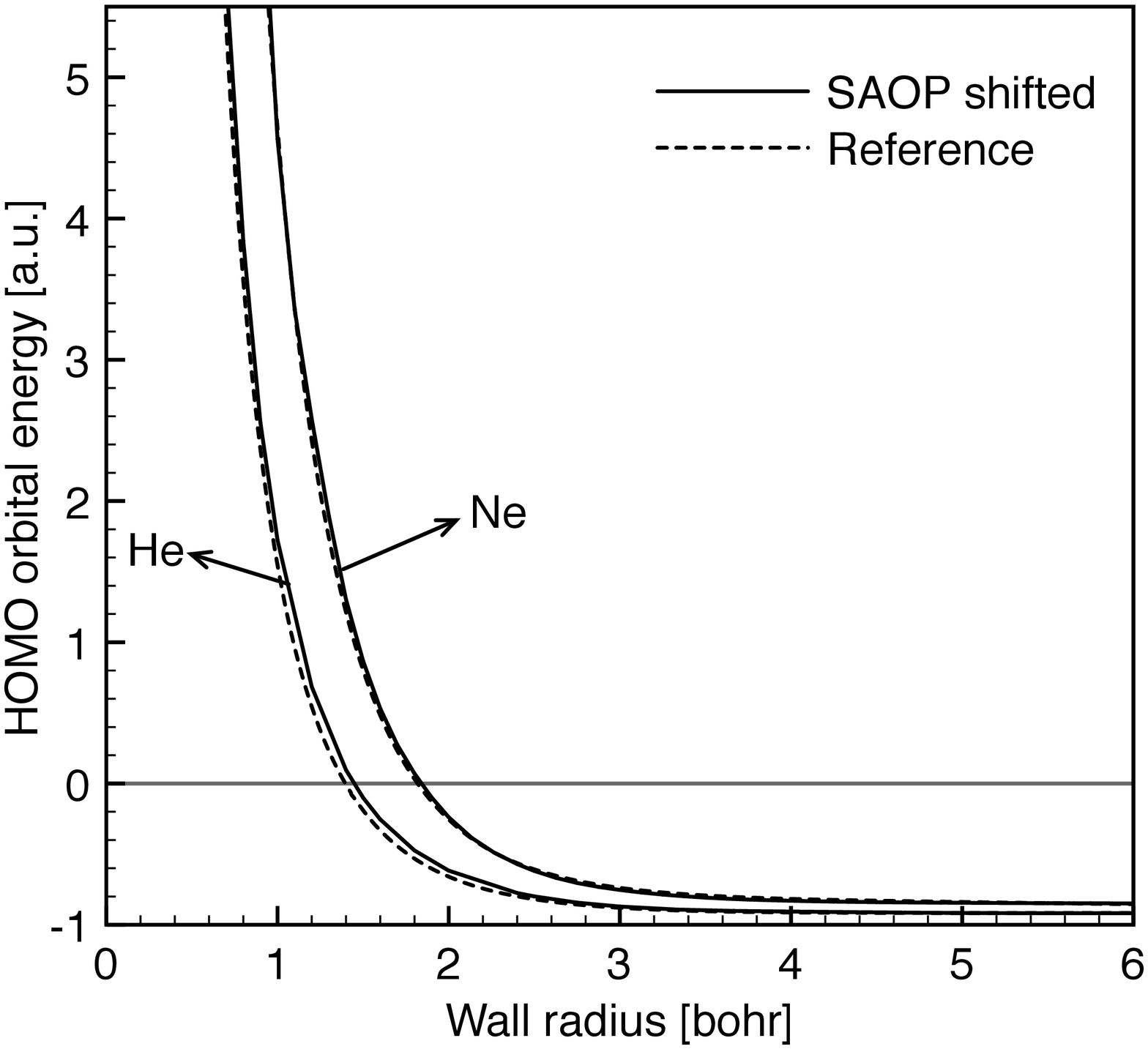}
\caption{\label{f:figure3} HOMO orbital energies of He and Ne as a function of the wall radii, compared with HF results from Ref.~\cite{L78}.}
\end{figure}
In the region where the HOMO energies are practically leveled off, i.e. after 4 bohr, the difference between the HF and SAOP values is almost constant.  For He we find a free atom value of -0.8165 a.u., the HF reference value is -0.9174. For Ne our SAOP free atom energy is -0.7585 a.u. and the reference HF value is -0.8511 a.u.. The differences are a result of the difference between the HF potential and our SAOP potential. The free atom value of He is equal to the HF value when the exact xc-potential is used, but as mentioned in the previous section, this potential is not valid when the atom is enclosed in a box of finite radius. We can try to correct for the shift by vertically shifting the SAOP curves with the free atom energy difference. We show the result in Fig.~\ref{f:figure3}. The shifted He and Ne results are almost on top of the HF results.

The ionization radius is defined as the radius where the HOMO level of the confined atom jumps from a negative to a positive energy when the box becomes smaller. The ``ionized'' electrons are still confined to the box, and therefore not explicitly ionized as such. The atomic ionization radii are 1.50 bohr for He and 1.89 bohr for Ne, these values correspond well with the 1.41 bohr and 1.83 bohr of Ref.~\cite{L78}. In Ref.~\cite{GVAS05} a DFT result for the ionization radius of Ne was obtained with the PW potential using total energy differences, they obtained a value of 1.83 bohr. Our SAOP results are close to the Hartree-Fock (HF) values of Ref.~\cite{L78}. When we use the shifted SAOP curves to recalculate the ionization radii we obtain 1.45 bohr for He and 1.84 for Ne, very close to the HF values.  

\begin{figure}[tbp]
\includegraphics[width=8.5cm]{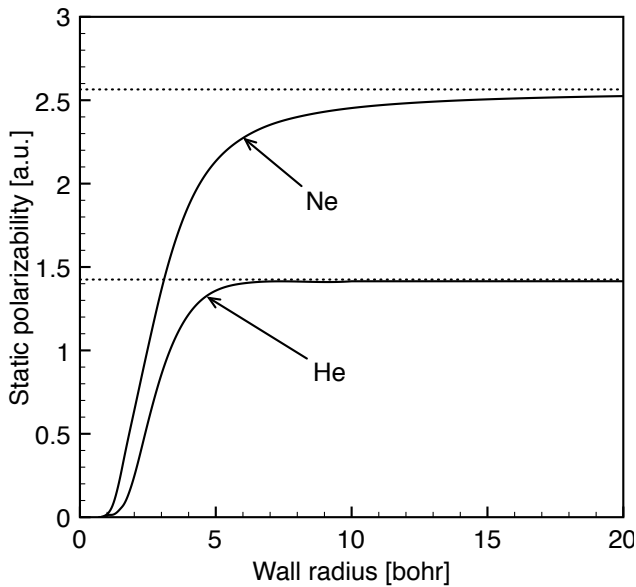}
\caption{\label{f:figure4} Static dipole polarizability of He and Ne as a function of the wall radii. The dotted horizontal lines indicate the respective free atom values.}
\end{figure}
In this work we present, to our knowledge, the first TDDFT polarizabilities for confined many electron atoms, and the first results including correlation. All previous works on many electron atoms used the HF approximation and obtained the polarizabilities using finite field methods. We use the SAOP potential in the ground state and the ALDA kernel for the response calculation of the polarizability. We show our results for He and Ne in Fig.~\ref{f:figure4}.
For small wall radii we see that the polarizability approaches zero. This is consistent with the findings that for one-electron systems, the results of which are expected to at least qualitatively apply to many-electron systems, the polarizability increases from $R=0$ as $\sim R^{4}$~\cite{F84}. For large radii the polarizability asymptotically approaches the free atom value from below as expected.  

The free atom values we predict are 1.425 $a_{0}^{3}$ for He and 2.565 $a_{0}^{3}$ for Ne, which correspond well with the experimental values of 1.41 $a_{0}^{3}$ for He and 2.57 $a_{0}^{3}$ for Ne~\cite{NIST_pol}.
We obtain an overall s-shaped curve, in agreement with the s-shaped curves found in Ref.~\cite{CSb03,CSc03} for several many electron atoms within the Hartree-Fock-Slater approximation. It is clear that the asymptotic approach of the Ne polarizability towards the infinite energy values is much slower than that of the He atom. As can be seen from Eq.~\ref{e:response}, the polarizability depends on all occupied and unoccupied states, including the continuum.  The wall always has an influence on the higher Rydberg states, since the potential of the neutral atoms is long-ranged. The (pseudo) continuum states in the box converge when the short-ranged part of the potential, that part that differs from $-1/r$, has converged. For He this point is reached well before 10 bohr, while for Ne this point is reached much later. This explains why the convergence of the Ne polarizability is so much slower than the He polarizability.

We expect the results for He and Ne to be representative for other closed shell atoms. For both the HOMO energies and the polarizabilities we reproduce the trends observed in other works. The BSTO basis set works well for these systems.

\subsection{Electron scattering: $e$-He${}^{+}$ and $e$-H}
\begin{figure}[tbp]
\includegraphics[width=8.5cm]{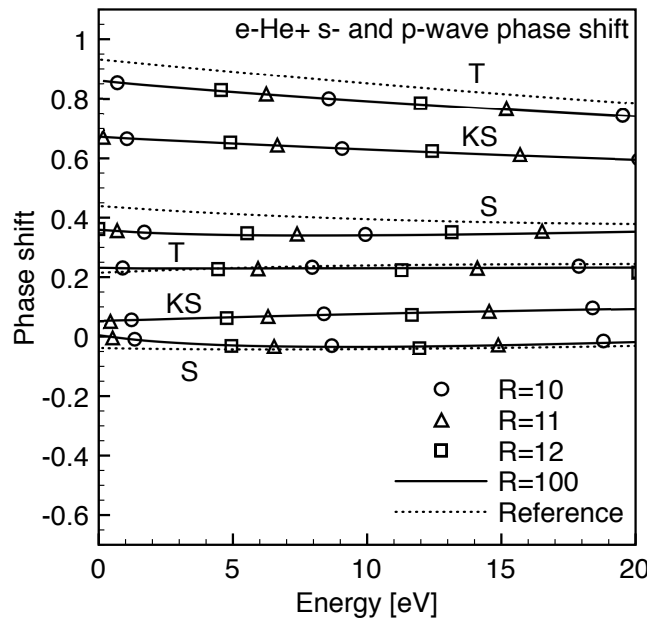}
\caption{e-He${}^{+}$ $s$- and $p$-wave scattering phase shifts (bare KS, singlet, and triplet) for BSTO basis sets with different wall radii. The bottom three curves correspond to $p$-wave scattering and the upper three curves to $s$-wave scattering. We compare the BSTO results to the phase shifts for a large radius of 100 bohr. Reference literature results are also included~\cite{B02,B06}.}
\label{f:figure5}
\end{figure}
\begin{figure}[tbp]
\includegraphics[width=8.5cm]{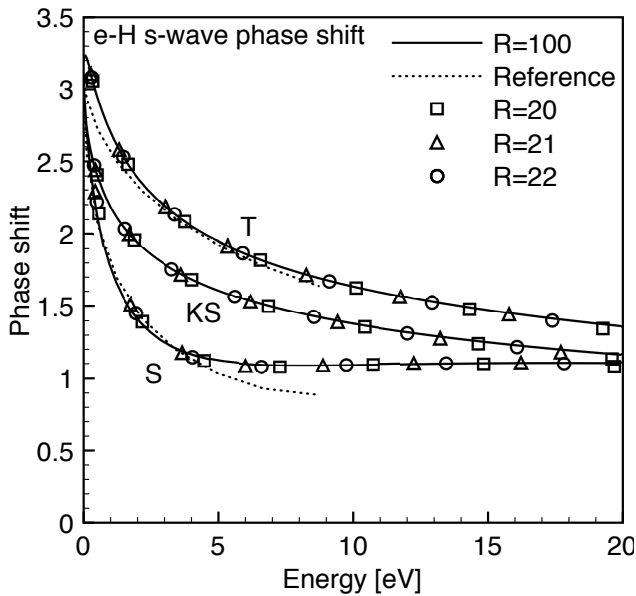}
\caption{e-H $s$-wave scattering phase shifts (bare KS, singlet, and triplet) for BSTO basis sets with different wall radii. We compare the BSTO results to the phase shifts for a large radius of 100 bohr. Reference literature results are also included~\cite{S61}.}
\label{f:figure6}
\end{figure}
We show results for the phase shifts corresponding to $e$-He${}^{+}$ and  $e$-H scattering in Figs.~\ref{f:figure5} and~\ref{f:figure6}. We compare our BSTO results with grid calculations corresponding to a wall radius of $R=100$ bohr.  The results of the different methods should coincide in case of a large basis set and a large grid. 

We first need to decide where to place the wall. We want to place the wall as close as possible to the nucleus to limit the necessary basis set size, but it needs to be in a region where the density and the occupied orbitals are negligibly small~\cite{FB08}. In case of He, the density has a range of approximately 2 bohr and the HOMO a range of approximately 6 bohr. We place the wall at $R=10,11,12$ bohr for this system. In case of H${}^{-}$, the density also has a range of approximately 2 bohr, but the HOMO orbital is very diffuse (its energy is only -1.257 eV) and it has a range of approximately 20 bohr. Therefore we place the wall at $R=20,21,22$ bohr. 

In Fig.~\ref{f:figure5} we see that results obtained with the BSTO basis set are indeed equal to the grid code results for e-He${}^{+}$ with $l=0,1$. Because we only have access to relatively small radii using the BSTO basis set, the points are far apart (the energies are widely spaced). To obtain the full curve, many different wall radii need too be used. Since the calculations are computationally very fast, this is in general not a problem. We see that even with only three different radii the curve is already well represented. On average 12 extra $s$- and $p$-functions are necessary on top of the standard ADF ET-pVQZ basis set  to obtain these results.

In Fig.~\ref{f:figure6} we show the BSTO results together with large radius results for $e$-H scattering with $l=0$. Since the wall distance is larger, we sample more points for each range. Again the entire curve is well reproduced with only 3 wall radii. Only 15 extra $s$-functions were necessary on top of the standard ADF ET-pVQZ basis set to describe the range of 0-22 bohr.

Overall, we observe that the phase shifts are well reproduced by the BSTO basis set method.

\section{Conclusions}
In this paper we placed an atom inside an infinite spherical potential by using a modified STO basis set multiplied by a cutoff function. We used (TD)DFT for the calculation of the HOMO energy and polarizability in case of confined atoms and for the calculation of electron-scattering phase shifts.

By placing the potential wall at different distances close to the atomic nucleus we studied the evolution of the HOMO energy and polarizability of the confined He and Ne atoms. The HOMO energy and polarizability obtained with (TD)DFT follow the same trends as available HF results~\cite{L78,CSb03,CSc03}.

By placing the wall at a large distance from the atomic nucleus, we discretized the continuum and obtained electron-scattering phase shifts for $e$-He${}^{+}$ and  $e$-H scattering. We showed that the same accuracy can be obtained with the basis set method as with a recently established grid method~\cite{FWEZ07,FB09}. The systems studied in this paper do not exhibit any low-energy resonances. When resonances are present the basis set method can be used in conjunction with the stabilization method~\cite{MRT93} to find their locations.

This research paves the way to study molecules in cavities and electron-molecule scattering using TDDFT. Going from the spherically symmetric system studied in this paper to molecules is not trivial. First we need to describe atoms that are off-center in a spherical cavity, this makes the evaluation of 1-center integrals more complicated. Such integrals have for example been evaluated for an off-center He atom in a spherical cavity using GTOs in Ref.~\cite{W99}. In case of molecules we also need to take 2-center integrals into account, a way to treat this problem using floating GTOs is described in Ref.~\cite{CS01}. Finally, the theory of electron-molecule scattering using TDDFT needs further development.

In conclusion, we showed that a unified basis set method can be applied to study both (time-dependent) properties of confined atoms and low-energy electron-atom scattering.

\section{Acknowledgments}

I acknowledge The Netherlands Organization for Scientific Research (NWO) for support through a Veni grant. I thank Pier Philipsen for useful discussions on the use of cutoff functions in the ADF BAND program. I thank Neepa Maitra for proofreading the manuscript. I also acknowledge the exact H$^-$ and He KS potentials from Cyrus Umrigar.

\end{document}